\documentclass{aa}
\usepackage{graphicx}
\usepackage{txfonts}
\begin{document}
   \title{Characterization of open cluster remnants 
   \thanks{Figures 3,7,9 and, 10 are only available in electronic form via http://www.edpsciences.org}}

   \author{D. B. Pavani
          \inst{1,}\inst{2}
          \and
          E. Bica\inst{2}
          }

   \offprints{D. B. Pavani}
 
\institute{Instituto de Astronomia, Geof\'isica e Ci\^encias Atmosf\'ericas (IAG)               		Universidade de S\~ao Paulo - Rua do Mat\~ao 1226; 05508-900, S\~ao                  	Paulo, SP; Brazil\\
		\email{daniela@astro.iag.usp.br}
	\and
Universidade Federal do Rio Grande do Sul, IF, CP 15051; 912501-970, Porto Alegre, RS; Brazil  \\
\email{bica@if.ufrgs.br}
            }

   \date{}

\abstract{}{}{}{}{} 
 
  \abstract
{Despite progress in the theoretical knowledge of  open cluster remnants and the growing search for observational identifications in recent years, open questions still remain. The methods used to analyze open cluster remnants and criteria to define them as physical systems are not homogeneous. In this work we present a systematic method for studying these objects that provides a view of their properties and allows their characterization.}
{Eighteen remnant candidates are analyzed by means of  photometric and proper motion data. These data provide information on objects and their fields. We  establish criteria for  characterizing  open cluster remnants, taking observational uncertainties into account.}
{2MASS J and H photometry is employed (i) to study  structural properties of the objects by means of  radial stellar density profiles, (ii) to test for any similarity between objects and fields with a statistical comparison method applied to the distributions of stars in the CMDs, and (iii) to obtain ages, reddening values, and distances from the CMD, taking an index of isochrone fit into account. The UCAC2 proper motions allowed an objective comparison between objects and large solid angle offset fields.}
{The objective analysis based  on the  present methods indicates 13 open-cluster remnants in the sample. Evidence of the presence of binary stars is found, as  expected for dynamically evolved systems. Finally, we  infer  possible evolutionary stages among remnants from the structure, proper motion, and CMD distributions. The low stellar statistics for individual objects is overcome by means of the construction of composite proper motion and CMD diagrams. The distributions of remnants in the composite diagrams resemble the single-star and unresolved binary star distributions  of open clusters. }
   {}


\maketitle
%

\section{Introduction}

The study of the space distribution, formation, age, and structure  of open clusters provides information on their 
evolution. Open clusters are destroyed over time by the action of both internal forces (mass loss through the dynamical 
evolution and stellar evolution) and  external ones, such as interactions with the Galactic tidal field, collisions
with molecular clouds, and/or disc shocking (Friel \cite{friel1999}; Wielen \cite{wielen}). Portegies Zwart et al. 
(\cite{portegies}) confirm central Galaxy tidal-field effects by showing that star clusters located inside 150 pc of the
Galactic center can be dissolved in $\sim50$ Myr.  In the solar neighborhood, most open clusters evaporate completely 
in less than 1 Gyr (Bergond et al. \cite{bergond}; Bonatto et al. \cite{bonatto2005a}). However, older objects exist (Lyng\aa~\cite{lynga}; Dutra \& Bica \cite{dutra}). By means of the open cluster catalogue by Dias et al. (\cite{dias}),
updated to January 2006\footnote{http://www.astro.iag.usp.br/$\sim$\,wilton}, we concluded that objects older than 1 Gyr amount to ${\approx}$ 10 \% of the 864 clusters with available ages.

Open cluster remnants (hereafter, OCR) correspond to final evolutionary stages. Numerical simulations (Terlevich \cite{terlevich}; de La Fuente Marcos \cite{fuente1997}, \cite{fuente1998}) have shown that the final stellar content of OCRs depends on their initial mass function, fraction of primordial binaries, Galactocentric distance, and total mass. The simulations also show that a strong dependence exists between the dissolution timescale of an open cluster and the stellar mass evaporation. As a consequence, OCRs are expected to be rich in binary stars. Owing to dynamical processes that dominate the dissolution, it is reasonable to assume that the effects of mass segregation will cause depletion of the low main sequence (MS)  in OCRs. Considering the above results with those of Hsu et al. (\cite{hsu}), Friel (\cite{friel1999}) made further developments towards an evolutionary scenario. Small and poorly populated open clusters dissolve primarily   from internal dynamical effects on the order of a  few 10$^{8}$ Myr, which explains most of the open cluster population. Apparently there is no  relationship between dissolution rate and  location in the Galaxy for low-mass open clusters. For richer and more massive open clusters,  the Galactic field effects become significant. Those with intermediate masses ($\sim 500$ to $1000M_{\odot}$)  can survive several Gyr, if they are located in the external regions of the  disc (e.g., NGC\,752 and NGC\,3680). Open clusters that survive longer  (e.g. NGC\,6791) must be more massive or else located exclusively in external regions of the disc (NGC\,188), or both (Friel \cite{friel2002}).

From an observational point of view, an OCR can be defined  as a poorly populated concentration of stars resulting from the dynamical evolution of a more massive system (Pavani et al. \cite{pavani2001}). Despite theoretical progress, and the growing number of observed candidates,  open questions still remain, such as: (i) Is there a preferential location in the Galaxy for OCRs? (ii) Do different evolutionary stages exist among remnants? (iii) Is it possible to define criteria for the characterization of OCRs? 

Studies in the literature reflect the debate on the observational evidence of OCRs. For instance, NGC\,6994 (M\,73) was interpreted as a possible remnant of an old open cluster  by means of  high-resolution spectra studies, radial velocities, atmospheric parameters, and proper motions (Bassino et al. \cite{bassino}).  On the other hand, Carraro (\cite{carraro2000}) and Odenkirchen \& Soubiran (\cite{odenkirchen})  interpret NGC\,6994 as a random fluctuation  of the stellar density using CCD photometry in B, V, and I bands and high-resolution spectra of the six brightest stars, respectively.

Bica et al. (\cite{bica2001}) present 34 possible open cluster remnants (POCR) that are located at relatively high galactic latitudes ($|b| > 15^{\circ}$) and are underpopulated with respect to the usual open clusters.  They show a significant  density contrast of brighter stars as compared to the Galactic field. Among the objects in that list, some have already been studied in detail it was concluded that NGC\,1252 is a remnant at $d_{\odot}=0.64$ kpc from the Sun, with an age of 3 $\pm$ 1 Gyr. NGC\,1901 is a physical  system dynamically comparable to the  Hyades, at $d_{\odot}=0.45$ kpc, and with an age of 0.6 $\pm$ 0.1 Gyr. Both were studied with B,V photometry, from the Tycho-2's  proper motions and comparisons of object CMDs with model predictions for the Galactic field (Pavani et al. \cite{pavani2001}). Carraro (\cite{carraro2002}) concludes that NGC\,7772 and NGC\,7036 are remnants. NGC\,7772 is located  at $d_{\odot}=1$ kpc, with 14 members and an age of 1.5 Gyr. NGC\,7036 has 17 members and is located at  $d_{\odot}=1$ kpc with an age of 3-4 Gyr. He used  UBVI photometry for  color-color diagrams, CMD, and a radial stellar density profile. NGC\,1663 was found to be a remnant of 2 Gyr located  at $d_{\odot}= 0.7$ kpc (Baume et al. \cite{baume}). They employed high-resolution spectra of the brightest stars and 2MASS photometry (Skrutskie et al. \cite{skrutskie})\footnote{Two Micron All Sky Survey, All Sky data release, available at http://www.ipac.caltech.edu/2mass/releases/allsky},  together with Tycho-2's astrometric data. Pavani et al. (\cite{pavani2003}) find that Ruprecht\,3 is a remnant at  $d_{\odot}=0.72$ kpc with an age of 1.5 $\pm$ 0.5 Gyr, using J and H 2MASS photometry, integrated and individual star spectroscopy,  and statistical methods for subtracting field stars from the CMD. Bonatto et al. (\cite{bonatto2004})  presented NGC\,2180 as a probable missing link between  evolved open clusters and remnants using J and H 2MASS photometry. They discuss  luminosity and mass functions,  structure, and CMDs. They obtained  an age of $700\pm 70$ Myr, $d_{\odot}=0.91\pm0.08$ kpc, and observed  mass $m_{obs}\,\sim\,47M_{\odot} $. Villanova et al. (\cite{villanova}) obtained CCD UBVI photometry and medium/high resolution spectroscopy for  NGC\,5385, NGC\,2664 and Collinder\,21 and their fields. The analysis using star counts, photometry, radial velocity distribution, and Tycho-2's proper motions indicated they are non physical objects. Carraro et al. (\cite{carraro2005}) discuss the nature of 11 possible OCRs  by using the Southern Proper Motion (SPM) Program 3  combined with 2MASS photometry. They conclude that  ESO\,282\,SC26 is a probable physical group. 

The methods employed in the series of studies above are not homogeneous. The present study intends to characterize OCRs by using  a systematic  set of observational methods applied to a large sample of candidates. We include new POCRs and objects in common with previous studies.  This is essential for establishing criteria for the observational identification  and determination of OCR properties, taking observational uncertainties into account. 

In Sect.\,2 we present the criteria for object selection and their structural properties. In Sect.\,3 we describe a statistical comparison  test between distributions of stars in the CMD. In Sect.\,4 we discuss the J $\times$ (J - H) CMDs and derive ages, reddening values, and distances, together with a classification of the objects based on an index of isochrone fitting. In Sect.\,5 we present the proper motion analysis. In Sect.\,6 we characterize the OCRs. Finally, concluding remarks are given in Sect.\,7. 


\section{Methods}
\subsection{Object selection }
The selection of POCRs was based on poorly populated stellar concentrations included in the catalogues by Alter et al. (\cite{alter}) and Lyng\aa\, (\cite{lynga}). The objects have a significant stellar density contrast  as compared to the background. Among the initially selected objects, we kept those presenting evidence of evolutionary sequences in the CMD. We favored higher Galactic latitudes  to avoid contamination by disc stars. The final sample contains 17 objects previously described as star clusters in open cluster catalogues and a new one (Object 1), which was found on a sky survey plate by one of us (E. B.). Based on DSS\footnote{The Digitized Sky Survey available: http://cadcwww.dao.nrc.ca/cadcbin/getdss} and 2MASS data we determined accurate equatorial coordinates for the POCRs (Table \ref{table:1}).

Among the 18 sample objects, 8 were previously included in a list of 34 POCRs (Bica et al. \cite{bica2001}). Of these, 7 have been discussed in recent studies: (i) ESO\,425\,SC6 and ESO\,425\,SC15 were classified as non physical (Carraro et al. \cite{carraro2005}); (ii) NGC\,1663 was found to be  a remnant (Baume et al. \cite{baume}); (iii) Ruprecht 3, NGC\,1252 and NGC\,1901 were found to be  remnants (Pavani et al. \cite{pavani2003}, Pavani et al. \cite{pavani2001b}); (iv) NGC\,6994 (Bassino et al. \cite{bassino}; Carraro \cite{carraro2000}, Ordenkirchen \& Soubiran \cite{odenkirchen}) is discussed further in the present work.

\begin{table*}
\caption{POCRs and open cluster NGC3680}
\label{table:1}
\centering
\begin{tabular}{lcrrcccc}
\hline 
Name& T & l& b& $\alpha$& $\delta$& $R_{\rm{lim}}$& \\
&&($^{\circ}$)&($^{\circ}$)&(h\,\,:\,\,m\,\,:\,\,s)&($^{\circ}$\,\,:\,\,\arcmin\,\,:\,\,\arcsec)&(\arcmin)\\
\hline
\hline
NGC\,3680&OC&286.76&16.92&11:25:38&-43:14:30&$22.0\pm1$\\
\hline
\hline
NGC\,6481& C& 29.94&14.93&17:52:48& 04:10:00&$1.5\pm0.5$\\

M\,73, NGC\,6994& C&33.71&-39.93& 20:58:55&-12:38:03&$4.5\pm0.5$\\

NGC\,6863 &C&32.27&-17.99&20:05:07&-03:33:20&$1.5\pm0.5$\\

NGC\,1663, OCL-461& L&185.88&-19.66&04:49:18&13:09:47&$6.0\pm0.5$\\

ESO\,425\,SC6& L&235.39&-22.28&06:04:50&-29:10:59&$3.0\pm0.5$\\

ESO\,425\,SC15& L&236.37&-20.35& 06:14:35& -29:22:30&$3.0\pm0.5$\\

Ruprecht\,3,  OCL-642, ESO426SC33& C&238.77&-14.81&06:42:07&-29:27:15&$2.0\pm0.5$\\

ESO\,426\,SC26& L&239.62&-16.51&06:36:18&-30:51:30&$3.0\pm0.5$\\

ESO\,429\,SC2& L&242.59& -4.16&07:33:24&-28:11:15&$5.0\pm0.5$\\

Ruprecht\,31, OCL-711, ESO68SC9& C&250.08& -5.97& 07:42:58&-35:35:50&$2.0\pm0.5$\\

Waterloo\,6& C& 264.82& -2.67& 08:40:17& -46:07:55&$1.5\pm0.5$\\

ESO\,570\,SC12& L& 274.06& 35.89& 11:12:07& -21:21:00&$5.0\pm0.5$\\

NGC\,1252& L&274.08&-50.83&03:10:49&-57:46:00&$6.0\pm0.5$\\

NGC\,1901, Bok 1, OCL-791.1&L&279.03&-33.60&05:18:11&-68:27:00&$5.0\pm0.5$\\

Object\,1& C& 294.94&-0.63& 11:41:34&-62:25:05 &$2.0\pm0.5$\\

ESO\,132\,SC14&C&308.15& 0.20& 13:36:30&-62:12:49&$1.5\pm0.5$\\

ESO\,383\,SC10&L&312.13&27.08&13:31:30&-35:03:56&$3.0\pm0.5$\\
 
Lyng\aa\,8 , OCl-962, ESO 226SC1&C& 333.27& -0.07& 16:20:04& -50:13:59 &$1.5\pm0.5$\\
\hline
\end{tabular}
 \end{table*}

The intermediate-age open cluster NGC\,3680 is used as a comparison (see also Bica et al. \cite{bica2001}). It presents evidence of an evolved open cluster dynamical stage and resembles in many respects what is expected for OCRs. The central luminosity function of NGC3680 is nearly flat from the turnoff to the turnover (Bonatto et al. \cite{bonatto2004}). Depletion of the low MS in central parts and the presence of a corona rich in low-mass stars has been detected in NGC\,3680 (see also Anthony-Twarog et al. \cite{anthony})  The 2MASS data are a powerful input in the analyses, making possible the study of (i) structural properties, (ii) objects and fields by means of a statistical method, (iii) the CMD distribution to derive ages, reddening, and distances. 

The kinematical data were extracted from The Second U. S. Naval Observatory CCD Astrograph Catalog - UCAC2 (Zacharias et al. \cite{zacharias}). It allows  quantitative comparison  of the proper motions of the object and large solid-angle offset fields. The use of  complementary methods proved to be essential in the present study to deal with completeness effects affecting POCRs.

\subsection{Structural properties}
\label{structure}

The 18 objects present two characteristic structures in the DSS/XDSS images. The first one is poorly populated with small  angular sizes and is centrally concentrated. Such objects are hereafter referred to as compact (C in Table \ref{table:1}). The second structure type is more populated in stars, with larger angular sizes and  more scattered stars. We refer to them as loose (L in Table \ref{table:1}). 

\begin{figure*}
\centering 
\includegraphics[width=14cm]{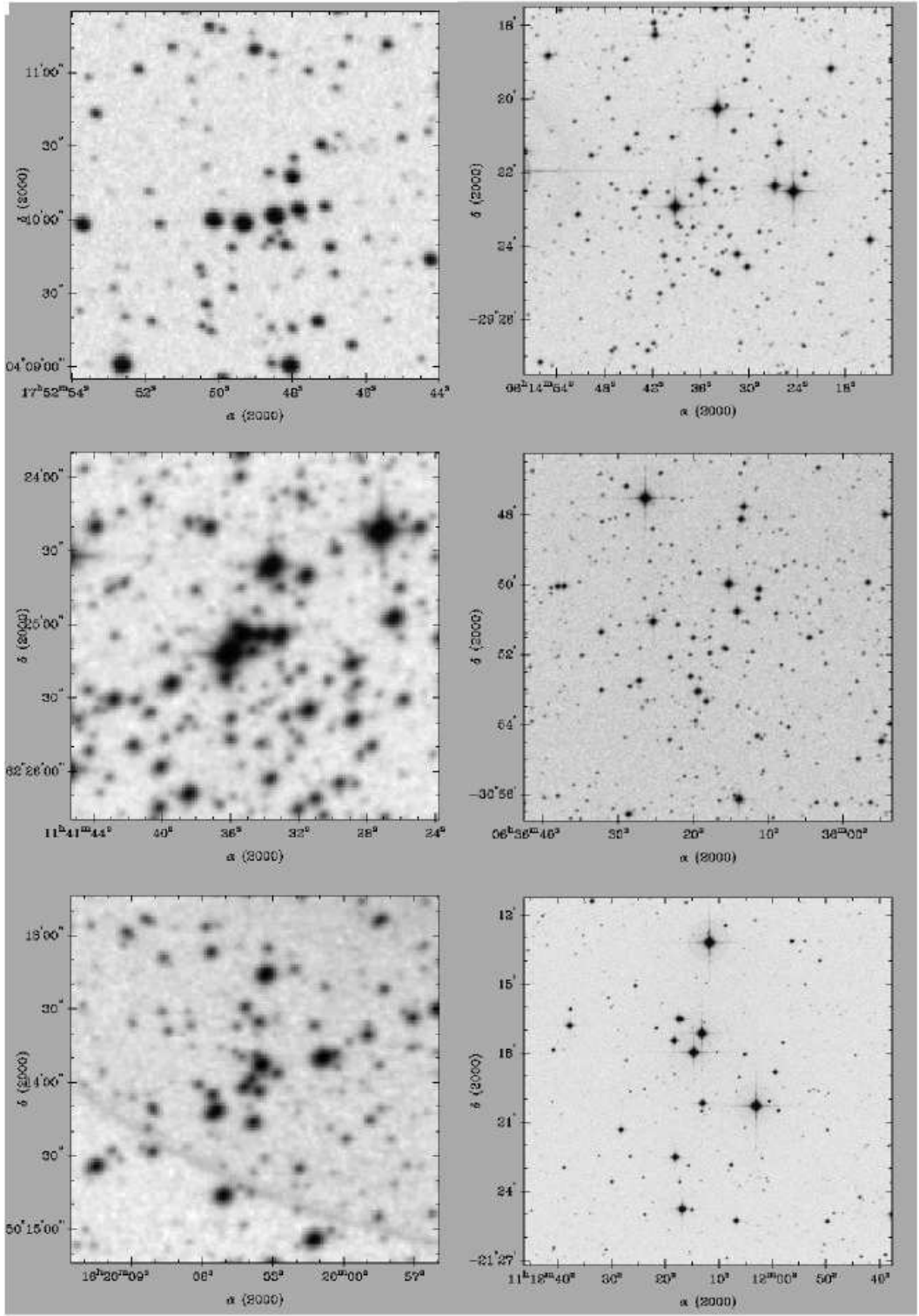}
\caption{XDSS R images for 6 sample objects. Left panels: compact
objects NGC\,6481 (top), Object\,1 (middle), and Lyng\aa\,8  (bottom). Right panels:
loose objects (downwards) ESO\,425\,SC15, ESO\,426\,SC26, and ESO\,570\,SC12. North is to the top and east to the left.}
	\label{FigXDSS}%
\end{figure*}

\begin{figure*}
  \centering
  \includegraphics[width=14cm]{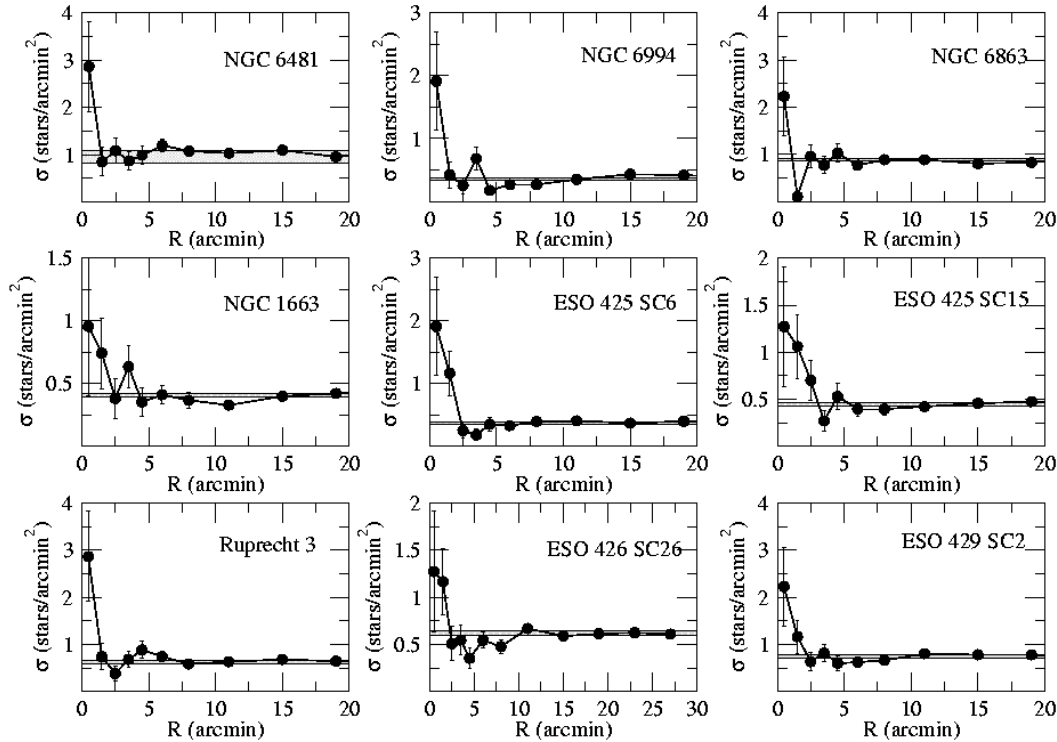}
\caption{Radial stellar density profile of sampled POCR. The average background level is shown (shaded rectangle). Poisson errors are also shown. For loose objects the dashed lines show the two-parameter King model fitted to the observed profiles.}
  \label{perfis}
\end{figure*}

\begin{figure*}
\centering
\includegraphics[width=14cm]{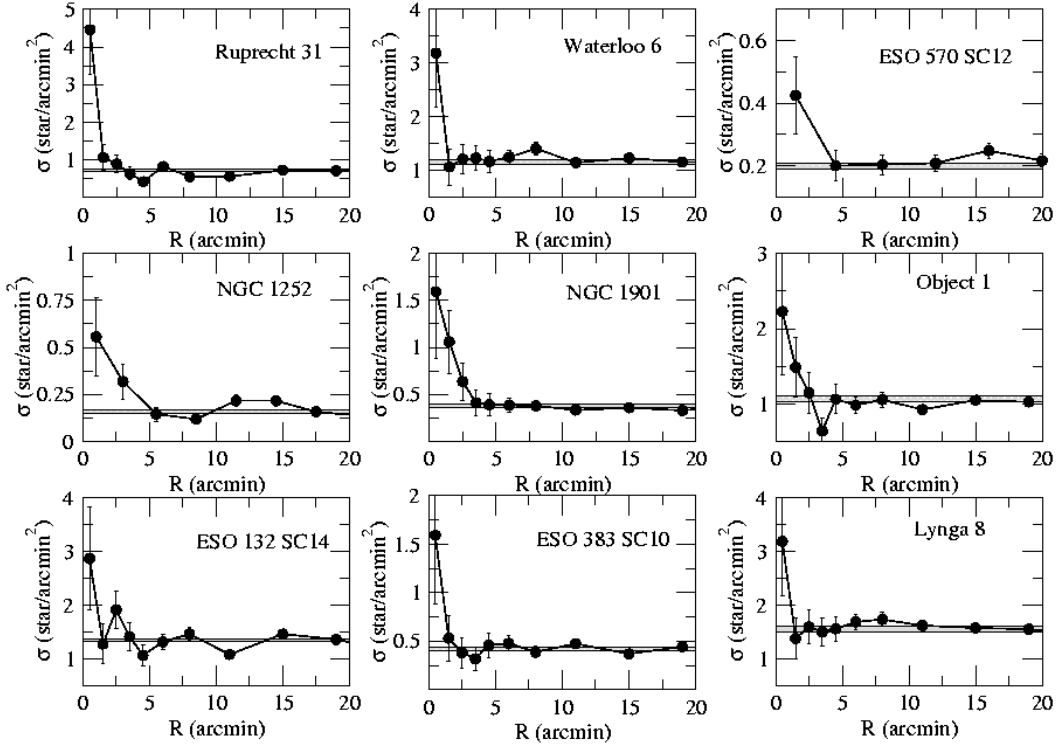}
\caption{Same as Fig. \ref{perfis}. }
\label{perfis2}
\end{figure*}

NGC\,3680 is similar to  loose POCRs in the present sample (Fig. 2, in Bonatto et al. \cite{bonatto2004}). Figure \ref{FigXDSS} shows examples of R band XDSS images of objects  classified by eye as compact (left panels) and loose ones (right panels). The compact objects present  a high-density contrast of stars as compared to the background and typically include up to 10 stars.  Table \ref{table:1} shows the 18  objects of the present sample, together with NGC3680. The columns give: (1) designation; (2) structural type (C for compact and L for loose); (3) and (4) Galactic longitude and latitude, respectively; (5) and (6) J2000.0 equatorial coordinates; (7) limiting radius $R_{\rm{lim}}$ obtained by means of the radial stellar density profile (Figs. \ref{perfis} and \ref{perfis2}).

The occurrence of two distinct structures might be associated to different evolutionary phases and dynamical processes leading to a depletion of stars. Another possibility would be a different origin for compact and loose objects related, e.g., to formation conditions and position in the Galaxy.

The 2MASS Point Source Catalogue completeness limit\footnote{http://www.ipac.caltech.edu/2mass/releases/allsky/doc/sec6\_5a1.html} is $99\%$  for $J<15.8$ and $H<14.1$ at $|b|>30^{\circ}$. Basically, areas of incompleteness are at $\pm75^{\circ}$ from the Galactic center, $b=\pm1^{\circ}$ from the Galactic plane, and radius   $r<5^{\circ}$ of the Galactic center. For these regions
the completess limits are $ J<14.0$  and $H<13.5$ mag.
Figures 7 to 9 of the 2MASS analysis of Release
Catalogs\footnote{http://www.ipac.caltech.edu/2mass/releases/allsky/doc/sec6.html} allow estimation of completess as a function of l and b. For the POCRs in Table \ref{table:1} with $|b|>14^\circ$, the derived limits  are $ J<15.8$ and $H<14.1$ mag, and for those with $|b|<6^{\circ}$ $J<14.0$ and $H<13.5$ mag.
 Object 1, ESO132 SC14, and Lyng\aa\,8 
are close to the plane, and the CMD
contains very red stars, so we applied a color filter (Sects. \ref{CMD} and \ref{MP}) to eliminate them, as well as to eliminate the contamination of LMC stars in NGC\,1901 direction (see Fig \ref{cp2}).
Figures \ref{perfis} and \ref{perfis2} show the radial density profiles for the magnitude-limited extractions 
for the whole sample.

The radial stellar density profile provides a general view of an open cluster structure. We apply this method to POCRs. Star counts with 2MASS photometry are carried out in concentric rings to take POCR and field areas into account. The extractions were centered in  the equatorial coordinates of Table \ref{table:1}. The radial density profiles were obtained from star counts in concentric rings with steps in radius depending on  the object  properties, divided by the ring area. The background contribution level shown in Figs. \ref{perfis} and \ref{perfis2} as shaded rectangles corresponds to the average number of stars present in an outer ring. Poisson error  bars are significant since POCRs are intrinsically poorly populated.

\begin{table}
\caption{Statistical test for POCRs and the open cluster NGC 3680}
\label{table:2}
\centering
\begin{tabular}{lrr}
\hline 
Name& $R^{2}_{\rm{POCR}}$& $P$   \\
&(\arcmin) & (\%) \\
\hline
\hline
NGC3680 & 0.071 & $<1$\\
\hline
\hline
NGC\,6481  & 0.232 & $<1$ \\

NGC\,6994 & 0.006 & $<22$\\

NGC\,6863  & 0.104 & $<11.5$\\

NGC\,1663 & 0.017 & $<1$\\

ESO\,425\,SC6& 0.025 & $<1$\\

ESO\,425\,SC15 & 0.027 & $<1$\\

Ruprecht\,3& 0.077 & $<1$\\

ESO\,426\,SC26& 0.029& $<6$\\

ESO\,429\,SC2 & 0.025 & $<83$\\

Ruprecht\,31 & 0.197 & $<1.5$\\

Waterloo\,6 & 0.184 & $<20$\\

ESO\,570\,SC12& 0.005 & $<1$\\

NGC\,1252& 0.001 & $<1$\\

NGC\,1901 & 0.030 & $<1$\\

Object\,1 & 0.459 & $<1.7$\\

ESO\,132\,SC14& 0.380 & $<1$\\

ESO\,383\,SC10 & 0.024 & $<1$\\
 
Lyng\aa\,8  & 0.018 & $<1$\\
\hline
\end{tabular}
\end{table}

As comparison, Fig. 4 of Bonatto et al. (\cite{bonatto2004}) shows that the radial profile of the evolved open cluster NGC\,3680 is described by a King model with core radius  $r_c =2.3\pm0.41\arcmin$ and a limiting radius $R_{\rm{lim}} =22\pm1\arcmin$.  Objects in Figs. \ref{perfis}  and \ref{perfis2}  illustrate  behaviors in the present sample. Compact POCRs as a rule show only one point above the background density. It basically corresponds to diameters estimated on DSS and XDSS images. Halos are generally absent. King models could not be fitted to the profiles of most POCRs. Loose POCRs present a density profile with an extended  region (possibly a  halo) characterizing their limiting radius, as illustrated by  NGC\,1663 (Fig. \ref{perfis}). The limiting radius arises where an observed profile merges with the background. We show the resulting limiting radii in Table \ref{table:1}.
The loose POCRs NGC\,1663, ESO\,425\,SC6, ESO\,425\,SC15, ESO\,429\,SC2, ESO\,383\,SC10 and NGC\,1901 could be basically fitted by a King model. A comparison with the evolved open cluster NGC\,3680 (Bonatto et al. \cite{bonatto2004}) shows that  Poisson errors are larger for  POCRs.  It is important to explore loose POCRs as possibly intermediate evolutionary dynamical stages between open clusters and OCRs. Despite the loose appearance of ESO\,570\,SC12 (Fig. \ref{FigXDSS}), its profile resembles those of compact POCRs, which might also suggest an evolutionary transition.
 
We conclude that the radial stellar density profiles of the sample objects are basically  consistent with  OCRs that gradually lose stars as compared to open clusters, persisting in the more internal parts responsible  for observed limiting radii.

\section{$R^2$ statistical test}
\label{statistical}

Automated procedures for a systematic search of star clusters in catalogues have been developed in the past years. Reyl\'e \& Robin (\cite{reyle}) combine the density of stars and the integrated flux in the $K_{\textit{s}}$ band. Ivanov et al. (\cite{ivanov}) employed the apparent stellar surface density, $K_{\textit{s}}$ band luminosity function and the distribution of stars in $J - K_{\textit{s}}$.   Mercer et al. (\cite{mercer}) present an  algorithm that uses statistical methods to locate high stellar densities aided by criteria of color and magnitude selections.  In general the main objective of those searches is to reveal any hidden globular clusters, infrared star clusters, and  young star clusters embedded in HII regions in our Galaxy. For known star clusters, there are e.g. studies from Mighell et al. (\cite{mighel96})  and Kerber \& Santiago  (\cite{kerber2005}) that applied statistical methods to compare populated clusters and fields using the CMD.

Although  OCRs are less populated than open clusters, they
should still present a stellar density contrast with respect to the surrounding field (Sect. \ref{structure}). However, it is necessary to adapt the open cluster comparison methods owing to the  intrinsically poorly populated nature of OCRs.  Bica et al. (\cite{bica2001}) compared the density of the bright  stars in the POCR area with what is predicted for the field from a model of Galactic structure and  GSC\footnote{The HST Guide Star Catalog, Version 1.2, available at http:// www-gsss.stsci.edu/} (Lasker et al. \cite{gsc}) counts in an offset equal area.
 
We further develop and adapt statistical comparison methods to the 2MASS data of POCRs and test the hypothesis that the object can be reproduced in radial stellar density and  CMD distribution by an equal-area  field fluctuation. To perform this  we employ a nearby large-area offset  field, with typical angular radius of $30\arcmin$ and separated from the object by $R_{\rm{lim}} + 3\times R_{\rm{lim}}$. Typically more than 100 sampling field areas of size equal to the $R_{\rm{lim}}$ of each  object  (Table \ref{table:1}) were generated.  Each  CMD was divided into boxes with dimensions in  color $\bigtriangleup{(J-H)}=0.125$ and magnitude $\bigtriangleup{J}=0.5$ mag, generating a grid. We considered boxes containing at least one star. This grid was applied to the object and all samplings in the offset field. In each CMD box, the number of stars was counted and stellar densities were calculated using the object area.

Based on the $s^{2}$ (Kerber et al. \cite{kerber2001}), we defined the $R^2$ statistics, which  reflects the distribution of field fluctuations and density contrasts  in the CMD plane between field and POCRs:

\begin{equation}
R^{2}=\sum_{i=1}^{n_{\rm{box}}}(\rho_{i}-\rho_{i,\rm{field}})^{2}
\end{equation}

\noindent where $n_{\rm{box}}$ is  the number of boxes in each grid for POCR (or sampled field CMD), $\rho_{i}$ corresponds to the box density for each POCR (or sampled field). Finally,  $\rho_{i,\rm{field}}$ corresponds to the box density  for the total field.

\begin{figure*}
 \centering
\includegraphics[width=14cm]{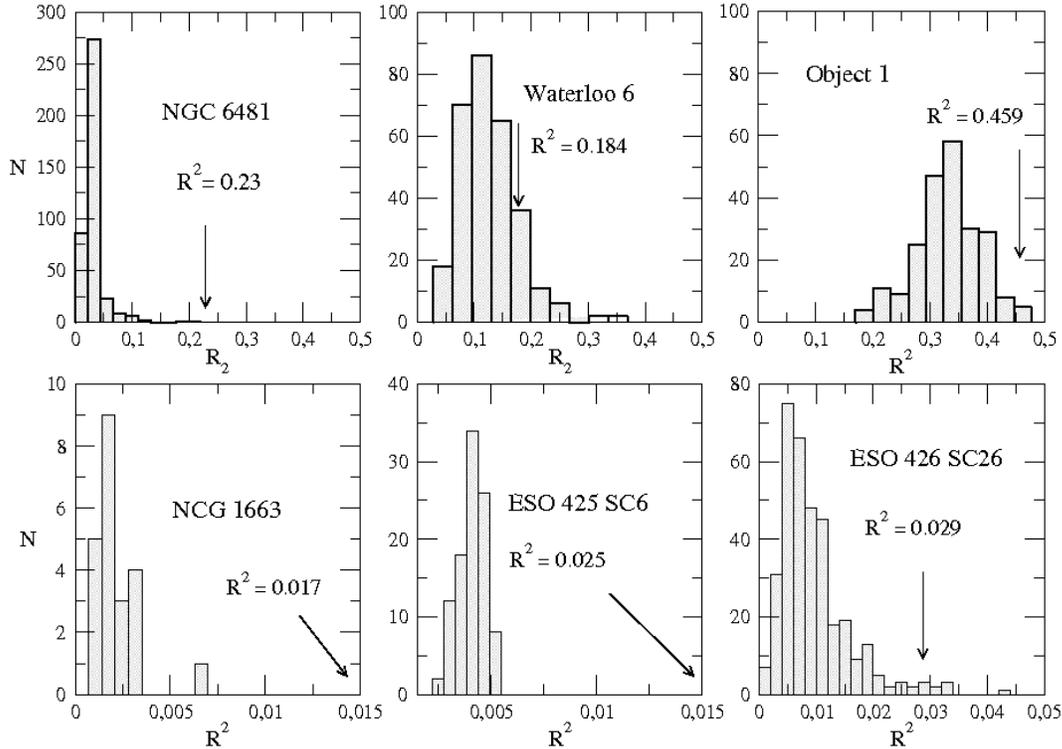}
\caption{Examples of $R^2$ statistical method testing  how similar CMDs of objects and respective fields are. Histograms of representative field regions within the $R^2$ interval are given. The arrow shows the object locus. Top panels: NGC\,6481 and Object\,1 are examples of compact objects and appear to be  physical ones, while  Waterloo\,6 appears to be a field fluctuation. Bottom panels: NGC\,1663, ESO\,425\,SC6, ESO\,426\,SC26 are examples of loose objects that appear to be  physical ones.}
         \label{R2}
   \end{figure*}

The ensemble of $R^{2}$ for the sampled fields (e.g. Fig. \ref{R2}) is a measure of the distribution of stars in the CMD of the total field, which in turn can be compared to that of the POCR. The method proved to be efficient for minimizing undersampling, which would otherwise be important if we had adopted the classical comparison between a populous open cluster and equal-area offset field.

Let $j_{\rm{POCR}}$ be the position of the POCR $R^{2}$ value 
($R^{2}_{\rm{POCR}}$) on the $R^2$ distribution for  $N_{\rm{s}}$ sampled fields; we then define the probability $P$ of a POCR  being a field fluctuation as:

\begin{equation}
P=1-\frac{|j_{\rm{m}}-j_{\rm{POCR}}|}{N_{\rm{s}}/2}
\end{equation}

\noindent where $j_{\rm{m}}$ is the median position in that distribution.
Table \ref{table:2} gives the $P$ and $R^2_{\rm{POCR}}$ values for the POCRs. 
If the $R^{2}_{\rm{POCR}}$ value falls  outside the $R^2$ field distribution, we adopt $P<1\%$. NGC\,6994,  Waterloo\,6, and ESO\,429\,SC2 attain $P<22\%$, $P<20\%$, and $P<83\%$, respectively, and appear  to be field fluctuations. NGC\,6863 presents $P<11.5\%$, and the remaining POCRs in Table \ref{table:2} have very low $P$ values, favoring physical systems.

\section{CMD Analyses}
\label{CMD}

\begin{figure}
\resizebox{\hsize}{!}{\includegraphics{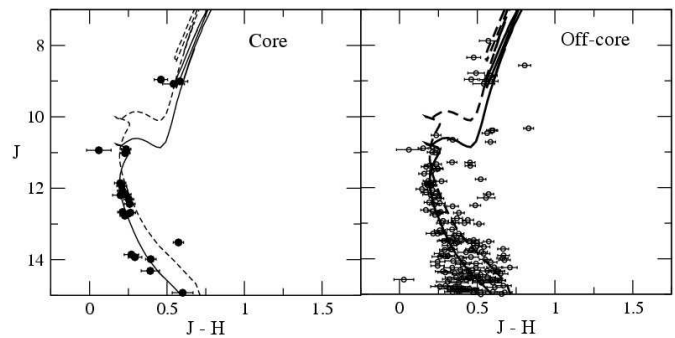}}
\caption{J$\times$(J - H) CMD of NGC\,3680 in core (left panel) and off-core (right panel) regions. Padova isochrones with solar metallicity and  shifted ones according to binarity (dashed line) are shown. Photometric errors are indicated.}
\label{n3680_cmd}%
   \end{figure}

We built 2MASS CMDs of  POCRs and offset fields with equal extraction radii, corresponding to  the region above the background field contributions in radial stellar  density profile (Figs. \ref{perfis} and \ref{perfis2}). We applied a cutoff at $J=15$ and excluded sources with flags for non-stellar objects for all CMDs  in order to minimize the star field contamination in POCR directions.  We derived reddening, distance modulus, distance to the Sun, Galactocentric distance, and age (Table \ref{table:3}) with the help of solar-metallicity Padova isochrones. In Fig. \ref{n3680_cmd} we present the open cluster NGC\,3680  core and off-core CMDs.  The core CMD (left panel) presents a star distribution that is expected for OCRs: depletion of the low MS in central parts and the presence of a corona rich in low-mass stars. The field of NGC\,3680 at $b\sim17^{\circ}$ is uncrowded according to the 2MASS $99\%$ completeness limit. NGC\,3680 is probably not affected by severe crowding, since the brighter limit found in 2MASS crowded fields\footnote{http://www.ipac.caltech.edu/2mass/releases/allsky/doc/sec6.html} occurs only near the plane or central parts of globular clusters. The central MS depletion (Fig. \ref{n3680_cmd}) was also observed by Anthony-Twarog et al. (\cite{anthony}). Since POCRs are less populous than open clusters, incompleteness is not expected. The brighter stars in the images, which in general characterize a POCR (Fig. \ref{FigXDSS}), were identified in the CMD,  which provided constraints for isochrone fittings.

\begin{figure*}
 \centering 
\includegraphics[width=14cm]{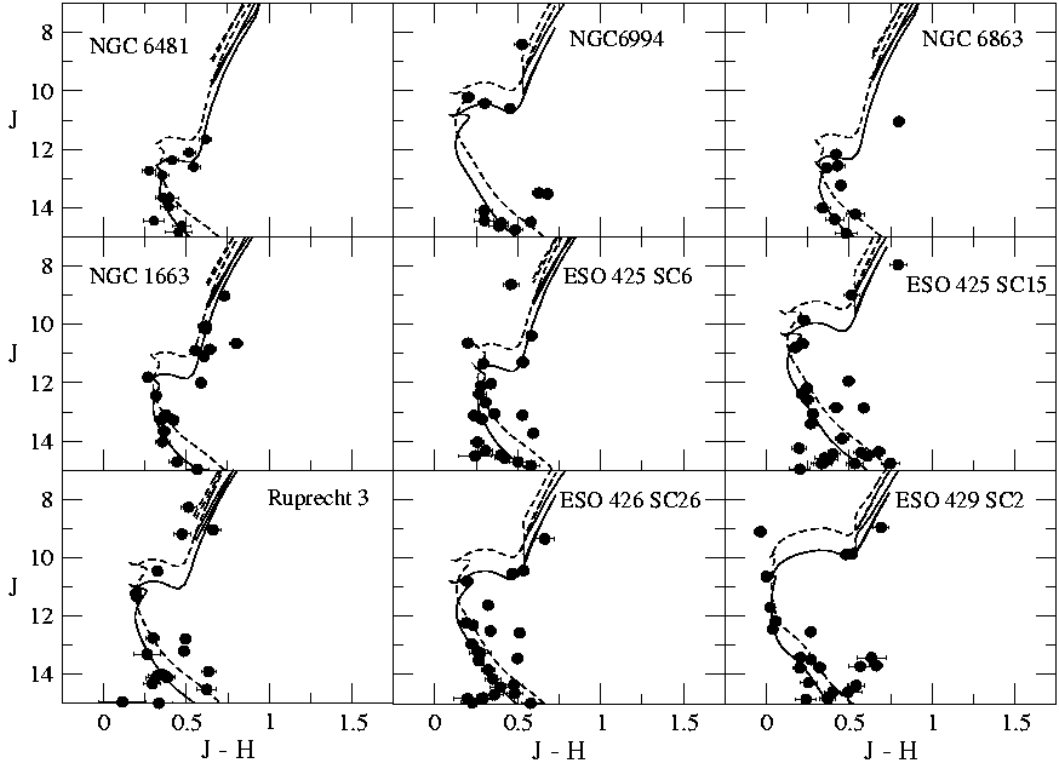}
\caption{J$\times$(J - H) CMDs of POCRs. Padova isochrones with solar metallicity, and the same  shifted according to binarity (dashed) are shown, and photometric errors indicated.}
\label{cmd_1}%
   \end{figure*}

\begin{figure*}
  \centering 
  \includegraphics[width=14cm]{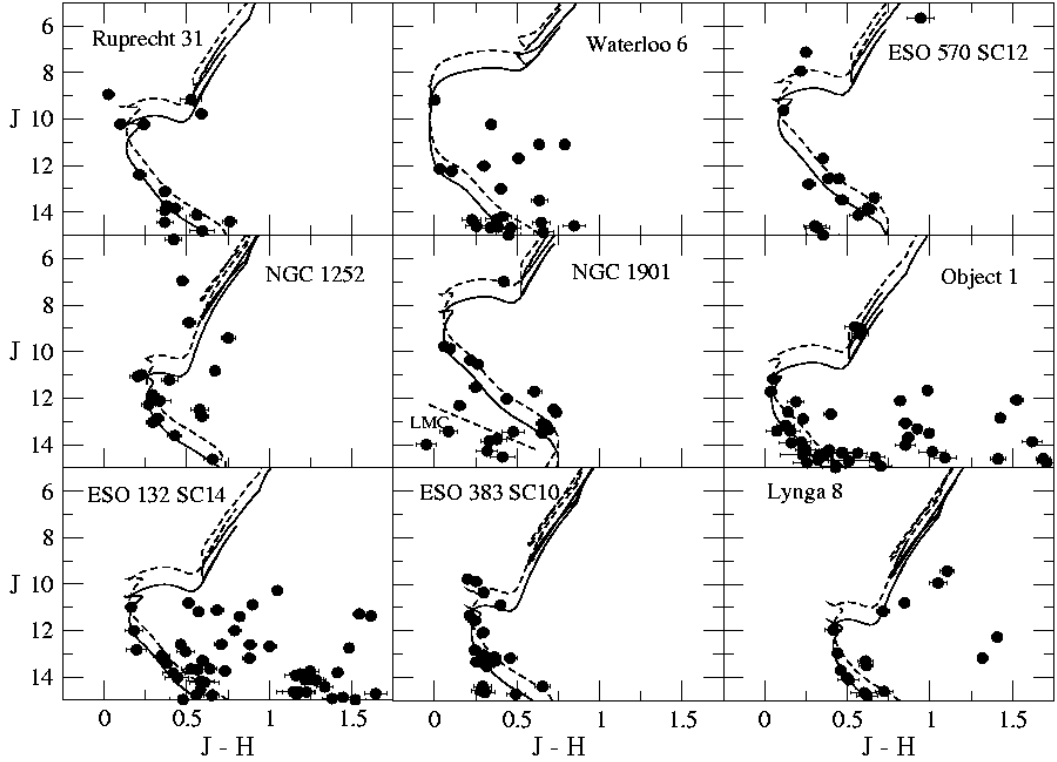}
  \caption{Same as Fig. \ref{cmd_1} for additional POCRs.}
             \label{cmd_2}%
	     \end{figure*}

\begin{figure*}
  \centering 
  \includegraphics[width=14cm]{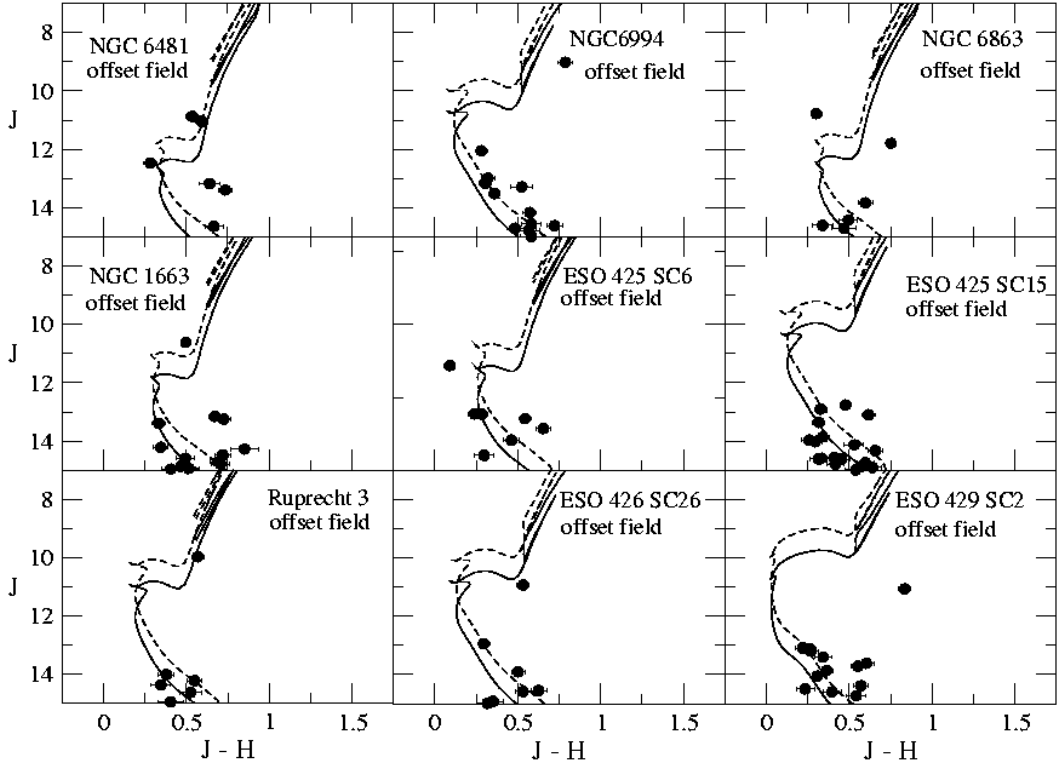}
  \caption{Same as Fig. \ref{cmd_1} for offset fields.}
             \label{cp1}%
	     \end{figure*}

\begin{figure*}
  \centering 
  \includegraphics[width=14cm]{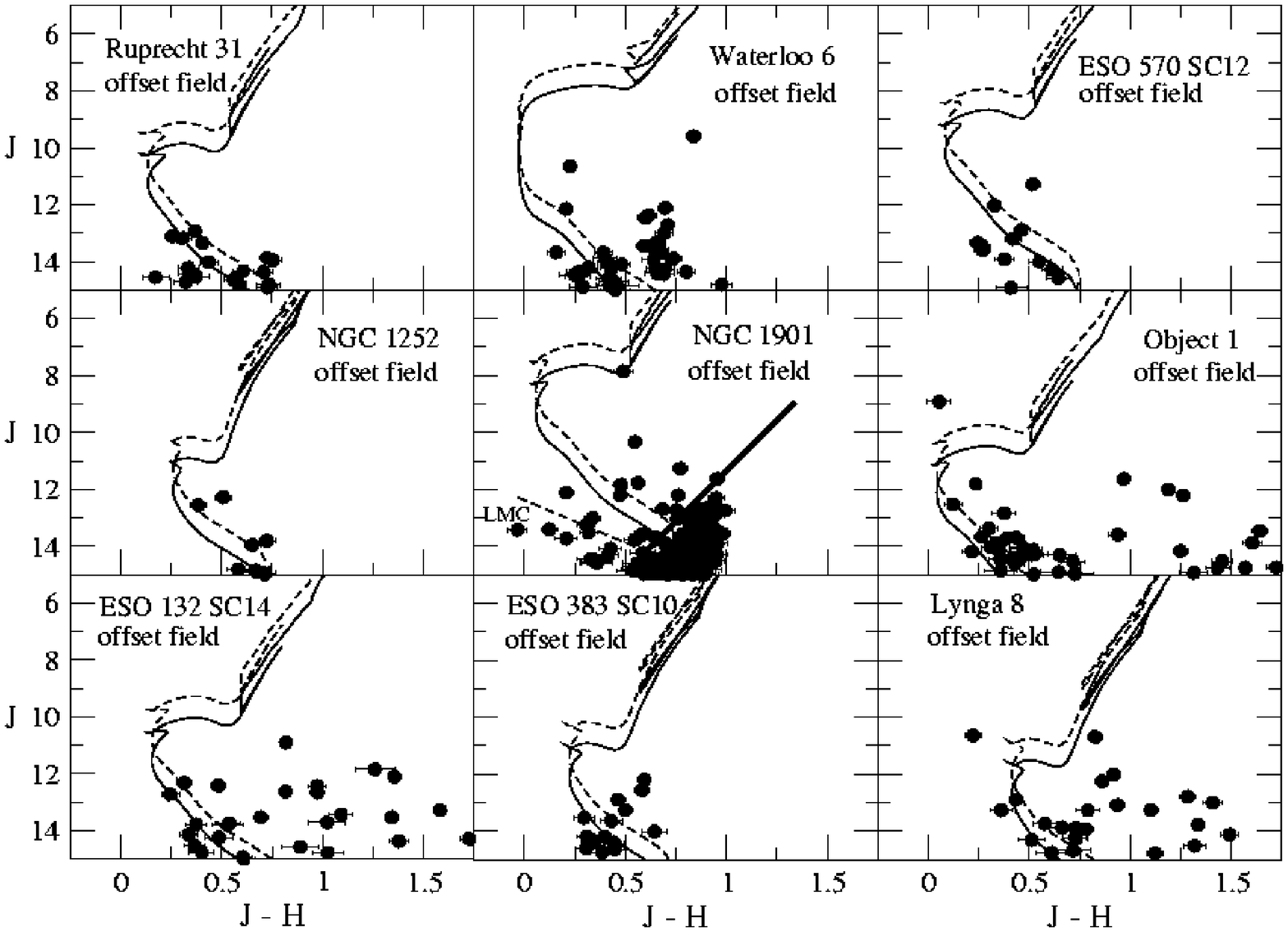}
\caption{Same as Fig. \ref{cmd_1} for additional offset fields. The thick line in the NGC\,1901 CMD indicates contamination of LMC stars eliminated by color filters.}
 \label{cp2}%
\end{figure*}

The NGC\,1901 direction is contaminated by LMC stars. We eliminated them by means of the color filters shown in Fig \ref{cp2}. To  verify  contamination in the NGC\,1901 direction we analyzed three  CMDs in three filters of fields with  the same angular size  as that of NGC\,1901: (i) the field of the globular cluster NGC\,1928 projected near the center of the LMC bar; (ii) the field of  the association of stars NGC\,2055 near the east end of the LMC bar dominated by a young population, and (iii) the field of NGC\,2004 with red supergiants.  Figures \ref{cmd_2} and \ref{cp2} dotted lines show the residual LMC contamination in NGC\,1901 and offset field directions, respectively.
Several POCRs have tight MS distributions, such as NGC\,6481, ESO\,425\,SC6, and ESO\,426\,SC26 (Fig. \ref{cmd_1}). A sequence of binaries, together with that of single stars in general, encompass most of the POCR stars. The sequence of binaries is obtained from that of single stars by a shift of $\bigtriangleup{J}=0.75$ mag (Santiago et al. 1996; Elson et al. \cite{elson}). As a rule, offset fields are underpopulated and, scattered and stars do not show the same brightness range (Figs. \ref{cp1} and \ref{cp2}). On the other hand, in the CMDs of Waterloo\,6  and ESO\,132\,SC14 (Fig. \ref{cmd_2}) stars are more scattered.

We applied a numerical method to quantify the fit of an isochrone to a POCR CMD. Stars compatible with single and binary isochrones were counted taking error bars into account. We obtained  their ratio to the total number of stars in the CMD, defined as  the fitting index $I_f$. Best isochrone fits could be obtained for all objects, and tentative fundamental parameters were derived (Table \ref{table:3}). This  suggests that only classical  CMD analyses may not be enough for conclusive results on the nature of  underpopulated  objects. The ensemble of methods will settle the more probable physical objects (Sect. \ref{discuss}). Table  \ref{table:3} gives (1) designation, (2) the extraction radius of the data, (3)  $I_f$, (4) age $\tau$ (Gyr), (5) near IR and (6)  optical reddening values, (7) distance modulus, (8) distance to the Sun $d_{\odot}$ (kpc), (9) linear diameter (pc), (10) height from the plane z (kpc), (11) Galactocentric distance $R_{gc}$ (kpc), and (12) number of probable members.

\begin{figure*}
  \centering 
\includegraphics[width=12cm]{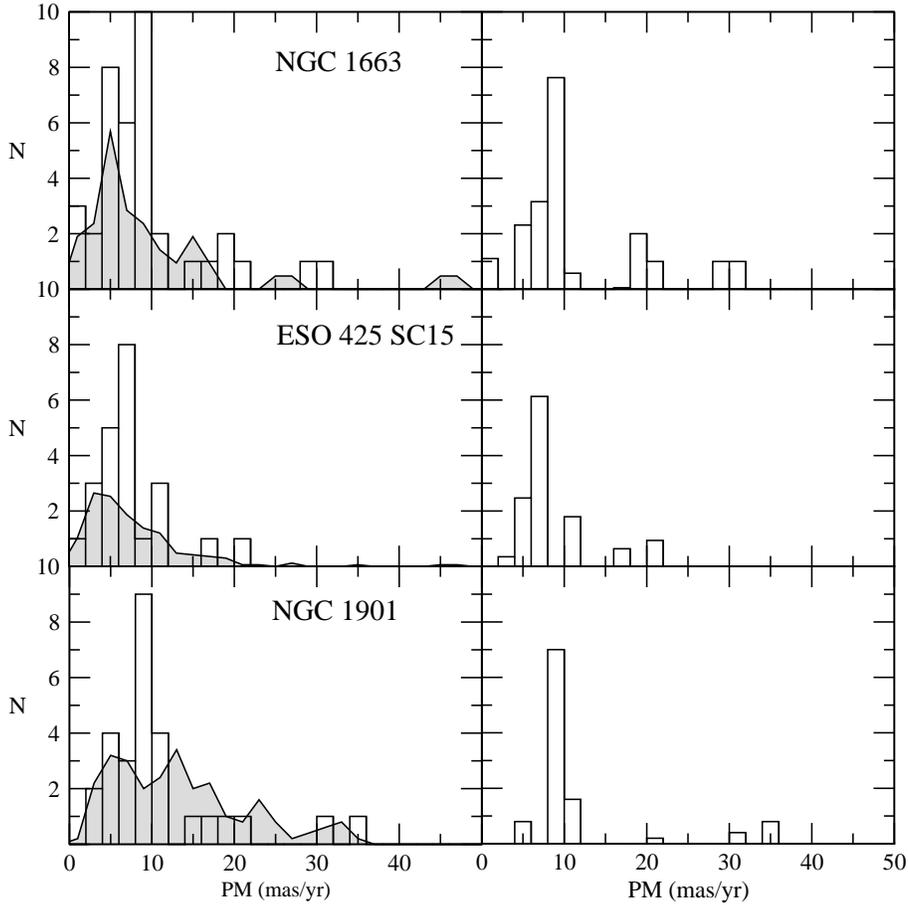}
\caption{Left panel: we illustrate the  modulus of the proper motion distribution  for the object region (histogram) and offset field (shaded area). Right panel: modulus of  the proper motion distribution after  field subtraction: these POCRs  are characterized by a well-defined peak.}
\label{grupoa_mp}%
\end{figure*}

\begin{figure*}
  \centering 
 \includegraphics[width=12cm]{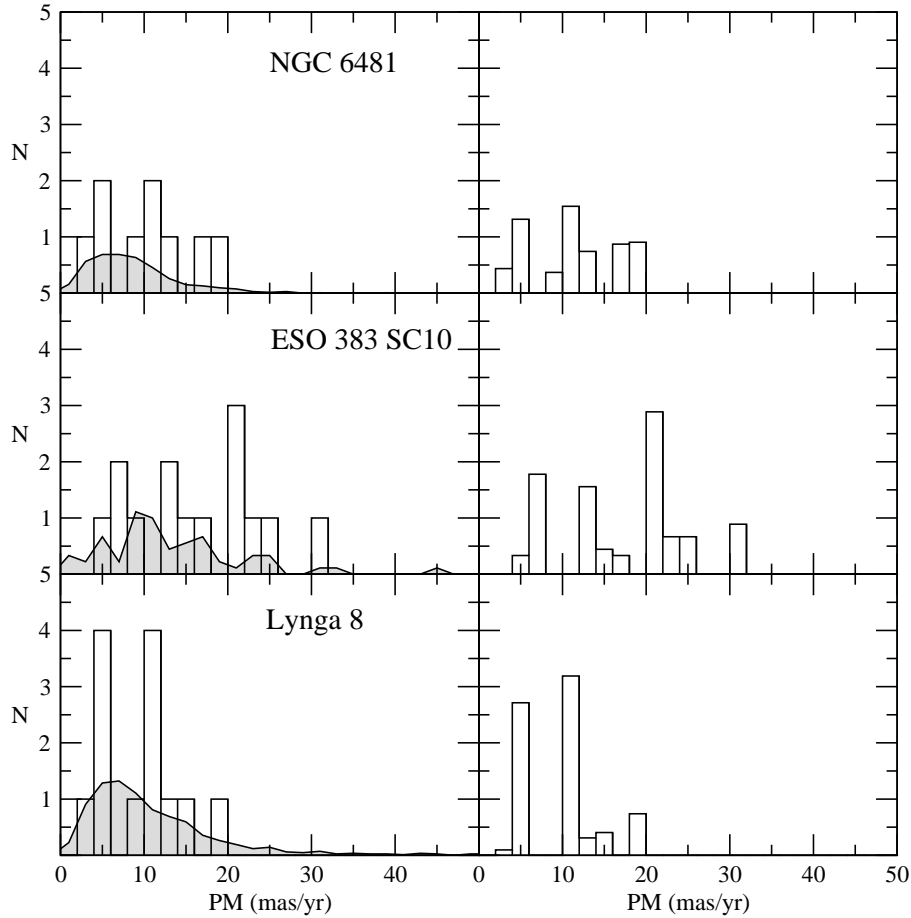}
  \caption{Same as Fig \ref{grupoa_mp}, for  POCRs  with a flatter distribution of the modulus of  proper motion. }
             \label{grupob_mp}%
 \end{figure*}

Linear diameters were are calculated from  angular limiting radii (Table \ref{table:1}) and  distances to the Sun. While  most POCRs have diameters  in the range 1-2 pc (Table \ref{table:3}),  open clusters such as M\,67 and NGC\,3680, at comparable ages and Galactocentric distances, have 17.4 and 12.8 pc, respectively (Bonatto et al. \cite{bonatto2004}). This suggests that POCRs correspond to open cluster cores. The E(B - V) values were coverted from E(J - H) using $E(B - V) = 3.03 \times E(J - H)$ (Rieke \& Lebofsky \cite{rieke}). The number of probable members was estimated from  $I_f$ as was the total number of stars in the POCR extraction radius. The number of probable members was not estimated for  POCRS with  $I_f<50\%$. Typical errors  are $\epsilon$(E(J - H)) = 0.02 mag, $\epsilon$(J - M$_J$) = 0.2 mag, $\epsilon$(d$_{\odot}$) = 0.1 kpc, $\epsilon$(z) = 0.015 kpc, $\epsilon$(R$_{gc}$) = 0.4 kpc. The errors combine 2MASS photometric errors with CMD fit errors. Using the relation $A_J/A_V = 0.282$ (Rieke \& Lebofsky \cite{rieke}), we derived $A_J = 2.65 \times E(J - H)$ in the calculations above. In the sample 13 POCRs with $I_f>60\%$ are described by the respective isochrones well. Three POCRs have  $50\% \leq I_f \leq 60\%$, while  Waterloo\,6 and ESO\,132\,SC14 have $I_f<50\%$.

We determined ages by fitting isochrones to the extracted CMDs and using equal area 
fields as references. No decontamination procedure was necessary. Most sample POCRs are located at relatively 
high Galactic latitudes and thus are favored by low contamination. The 6 POCRs at $|b|<6^{\circ}$
have small angular diameters and bluer colors on average than the field stars (Fig. \ref{cmd_2}).

\begin{table*}
\caption{CMDs parameters for POCRs and the open cluster NGC 3680}
\label{table:3}
$$
\centering
\begin{tabular}{lccccccccccc}
\hline 
Name& $r_d$& $I_f(\%)$& $\tau$ (Gyr)&  E(J - H) &$E(B - V)$&$(J - M_J)$& $d_{\odot}$& D& z& $R_{gc}$ &pm\\

& (\arcmin) & (\%)& (Gyr)&  (mag) &(mag)&(mag)& (kpc)& (pc)&  (kpc)& (kpc)\\
\hline
\hline
NGC\,3680& $3.0$& $78 \pm 18$& $1.6 \pm 0.01^{\mathrm{a}}$& 0.0$^{\mathrm{a}}$&0.0&10.00$^{\mathrm{a}}$&1.00$^{\mathrm{a}}$& 12.8$^{\mathrm{a}}$&0.30&8.8$^{\mathrm{a}}$&18\\
\hline
\hline
NGC\,6481 & $1.5$ &  $83 \pm 26$ & $3.5 \pm 0.5$& 0.06&0.18& 10.36 &1.18&1.03&0.30&7.04&10\\

NGC\,6994& $2.5$ & $83\pm 26$&$1.0 \pm 0.2$& 0.01&0.03 &9.77&0.9&2.36&-0.57&8.0&10\\

NGC\,6863 & $1.5$ &  $75 \pm 31$ & $3.5 \pm 0.5$&0.00& 0.00&10.40& 1.20&1.05&-0.37&7.06&7\\

NGC\,1663& $3.0$ &  $88 \pm 23$& $2.8 \pm 0.2$&0.04&0.12&0.20& 1.10&4.20&-0.40&9.13&15\\

ESO\,425\,SC6& $2.5$ &  $71 \pm 18$& $2.5 \pm 0.3$& 0.00&0.00&10.20&1.10&1.92&-0.41&8.62&15\\

ESO\,425\,SC15& $3.0$ & $61 \pm 15$ & $1.0 \pm 0.3$& 0.01&0.03&9.97&0.99&1.73&-0.34&8.55&15\\

Ruprecht\,3& $2.0$ & $65 \pm 18 $& $1.5 \pm 0.5$& 0.00&0.00&10.20& 1.10&1.28&-0.28&8.60&13\\

ESO\,426\,SC26& $2.5$ & $70 \pm 17$& $1.0 \pm 0.3$& 0.01&0.04&10.46&1.24&2.16&-0.35&8.66&17\\

ESO\,429\,SC2& $1.5$ & $68 \pm 17$ & $0.4 \pm 0.05$& 0.03&0.9&11.20&1.67&4.85&-0.12&8.88&15\\

Ruprecht\,31& $1.5$ &  $78 \pm 24$ & $1.0 \pm 0.3$& 0.02&0.06&9.85&0.93&0.81&-0.10&8.36&11\\

Waterloo\,6& $1.5$ & $48 \pm 15$& $0.2 \pm 0.1$& 0.00&0.00&10.5&1.26&1.28&-0.06&8.21&-\\

ESO\,570\,SC12& $4.0$ &  $62 \pm 20$& $0.8 \pm 0.2$&0.01&0.03&9.02&0.64&1.86&0.37&7.99&10\\

NGC\,1252& $4.0$ & $65 \pm 19$&$2.8 \pm 0.3$& 0.00&0.00&9.5&0.79&2.76&-0.61&8.00&11\\

NGC\,1901& $5.0$ & $67 \pm 21$& $0.6 \pm 0.1$& 0.01&0.03&8.35&0.46&1.34&-0.26&7.95&10\\

Object\,1& $1.5$ & $50 \pm 11$ &$0.6 \pm 0.1$& 0.00&0.00&11.20&1.74&2.00&-0.02&7.44&20\\

ESO\,132\,SC14& $1.5$ & $33 \pm 9$&$0.8 \pm 0.2$& 0.00& 0.00&10.20&1.10&0.95&0.003&7.37&-\\

ESO\,383\,SC10& $3.0$ & $76 \pm 19$&$2.0 \pm 0.5$&0.00&0.00&10.00&1.00&1.74&0.45&7.45&15\\
 
Lyng\aa\,8  & $1.5$ & $56 \pm 19$&$2.0 \pm 0.5$& 0.19&0.57&10.11&1.05&0.92&-0.001&7.07&9\\
\hline
\end{tabular}
$$
\begin{list}{}{}
\item[$^{\mathrm{a}}$  Bonatto et al. (\cite{bonatto2004}).]
\end{list}
\end{table*}

\section{Proper motions}
\label{MP}

Bica \& Bonatto (\cite{bica2005}) present internal proper motion distributions that are related to binary stars in open clusters. They show that proper motions are a useful tool for identifying higher-velocity stars as unresolved binary cluster members and, as a consequence, for mapping and quantifying the binary fraction in the CMD.

Individual POCRs show low stellar statistics as compared to open clusters,
and composite POCR CMDs  might provide more insight into their nature. Proper motions extracted from UCAC2 allowed a comparison between POCR areas and large offset  fields. 

We applied a color filter (Bonatto et al. \cite{bonatto2005b}) to the CMD of a given POCR and offset field stars that removed most  background contamination. The color filter was used to separate the CMD of a POCR
from background contamination (most disc and/or bulge stars), leaving a residual contamination
that will be taken into account by means of histogram  subtractions.  In the following we employ the modulus of proper motion components
 
\begin{equation}
MP=\sqrt{(\mu_{\alpha}cos\,\delta)^{2}+\mu_{\delta}^{2}}.
\end{equation}

\noindent To be consistent with the  CMD analysis, we extracted the data for each POCR inside the limiting radius $R_{\rm{lim}}$ (Table \ref{table:1}). We  verified that the correspondence between   UCAC2 and 2MASS is nearly complete for $J\leqslant14.5$.

Histograms of stars in proper motion modulus bins of 2 mas/yr were built for  POCRs with $I_{f}\geqslant 50\%$ and their offset fields. The offset field was scaled to the POCR area. We  subtracted  the offset field histogram from that of the POCR region, and the result provides the proper motion distribution in Figs. \ref{grupoa_mp} and \ref{grupob_mp}. Asymmetries and peaks in the proper motion distributions may yield information on the internal kinematics, presence of binaries, and possible evolutionary stages.

In order to analyze  intrinsic properties, we transformed proper motions into linear velocities on the plane of the sky ($v_p$). The units mas/yr convert to km/s using the distances  in Table \ref{table:3} and 

\begin{equation}
v_p (km/s)=MP(mas/yr)*d(pc)*4.74\times10^{-3}.
\end{equation}

\noindent The objects were divided into two groups  (A and B) according to their $v_p$ distributions characterized by peaks or flatter distributions. The  observed median velocities  ($v_M$) were derived for rest velocity corrections (Table \ref{table:4}), and the median is suitable for non Gaussian distributions.

Finally, composite velocity histograms were built: (i) $v_M$  in each POCR histogram  was used to correct the diagram to  rest velocity, providing it in the form ($v_p - v_M$); (ii) Group A is characterized by a well-defined low-velocity peak and is formed by the loose POCRs  NGC\,1663, ESO\,425\,SC15, ESO\,426\,SC26, NGC\,1252, NGC\,1901, and the compact one Ruprecht\,31. They have ($v_M < 71$ km/s). Group B is characterized by  higher-velocities,  $75 \leqslant v_M < 140$ km/s, and is formed  by the compact objects NGC\,6481, NGC\,6863, Ruprecht\,3, Object\,1, Lyng\aa\,8 and the loose ones,  ESO\,425\,SC6, ESO\,570\,SC12, and   ESO\,383\,SC10.

Figure \ref{groupab}  presents the composite histograms  ($v_p - v_M$) for Groups A and B.  Group A shows a well-defined low-velocity peak.  Group B shows a double peak. Table \ref{table:4} gives the POCRs belonging to each group. It is outstanding that the  ($v_p - v_M$) composite histogram  reveals properties of POCRs only seen before in open clusters (Bica \& Bonatto \cite{bica2005}), which cannot be observed in individual POCRs owing to low statistics.  The double-peak distribution in Group B  resembles  that of single (low velocity) and unresolved binary stars (high velocity) as observed in off-core regions of M67  (e.g. Fig. 5  in Bica \& Bonatto \cite{bica2005}).

\begin{table}
\caption{The groups and median velocity values}
\label{table:4}
\centering
\begin{tabular}{llr}
\hline 
Name& Type & $v_M$ (km/s)\\
\hline
\hline
Group A& &\\
\hline
\hline
NGC\,1663& L&68\\

ESO\,425\,SC15& L& 47\\

ESO\,426\,SC26& L& 70\\

Ruprecht\,31& C& 44\\

NGC\,1252& L& 68\\

NGC\,1901& L& 26\\
\hline
\hline
Group B & &\\
\hline
\hline
NGC\,6481 &  C& 78\\

NGC\,6863 & C& 137 \\

ESO\,425\,SC6& L& 83\\

Ruprecht\,3& C&83\\

ESO\,570\,SC12& L& 109\\

Object\,1&C& 82\\

ESO\,383\,SC10& L & 113\\
 
Lyng\aa\,8& C&75 \\
\hline
\end{tabular}
\end{table}

Several objects were not included in the group analysis because of doubt about their nature or they are field fluctuations because the indices are indicating non physical objects (Table \ref{table:5}). Although ESO\,429\,SC2 has a proper motion distribution compatible with Group A and has $I_f=68\pm16\%$, it was not included in the composite histogram due to the  probability of being a field fluctuation (Table \ref{table:2}). The CMD of NGC\,6994 (Fig. \ref{cmd_1}) shows a gap of $\approx$3 mag. between bright and faint stars that is improbable in a turnoff (TO) to low MS distribution. The statistical method (Table \ref{table:2}) also favors  a non physical nature. These results point to NGC\,6994 being a field fluctuation, in agreement with Carraro (\cite{carraro2005}) and Odenkirchen \& Soubiran (\cite{odenkirchen}).

Composite open cluster CMDs  have been a tool employed  since Sandage (\cite{sandage}) and were  useful for  modeling  stellar evolution. In the present study we  applied this principle to OCRs and POCRs. By means of reddening and distance corrections, we obtained absolute CMDs, which were in turn combined to build  composite CMDs for  Groups A and B, improving the stellar statistics in the diagrams (Fig. \ref{gpab_cmd}). Considering the ($v_p - v_M$) distributions for Groups A and B, stars with velocities higher than 25 km/s are more probably related to unresolved binary stars,  based on  analysis of  the populous open cluster M\,67 (Bica \& Bonatto {\cite{bica2005}). We plotted those stars  in the  composite CMD and they tend to be located to the right of the MS of single stars, as expected (Fig. \ref{gpab_cmd}).

Group A (Fig. \ref{gpab_cmd}) shows the low-velocity stars  distributed along the MS for different ages, while the higher-velocity stars are  biased towards the right side of the isochrones. This is the locus encompassed by the color and magnitude limits expected for binary stars. In the  composite CMD of Group B (right panel)  a similar   behavior is seen, and  stars around the  3.5 Gyr isochrone suggest that   higher-velocity stars  tend to replace the single-star MS. This  loss of  single-stars to the field would tend to cause a higher fraction of binaries in such OCRs, in agreement with the results of N-body simulations (de La Fuente Marcos \cite{fuente1997}, \cite{fuente1998}).

The  composite diagrams of  Groups A and  B suggest  possible evolutionary stages among the studied remnants, where  Group B would be more dynamically evolved than  Group A, and in general  loose objects would be less evolved than compact ones.

\section{Discussion}\label{discuss}

The present set of  methods for the  study of  POCRs  provided  a  view  of their  properties and  criteria for their characterization. Of the 18 sample objects,  ages, reddening values, and distances could be determined for 16 of them having $I_f >50\%$ (Table \ref{table:3}). In Table \ref{table:5} we present a final POCR classification according to different criteria: (i) radial stellar density profile  considering  Poisson errors, (ii) probability of  less than $15\%$ for the POCR CMD to be  representative of the offset field CMD, (iii) homogeneous distribution of  stars in the CMD and fitting index $I_f \geq55\%$.  Columns give the partial classifications  in each method, where ``y'' stands for the attained minimum value, while  ``n'' stands for those not attained. The final results are indicated in the last column, where  ``OCR'' is an object we concluded is an open cluster remnant, FF   a  field fluctuation, and POCR is maintained as a possible open cluster remnant.

Structural  and CMD analyses  together are powerful tools for inferring the nature of POCRs. As an example, although NGC\,6994 has a reasonable density profile, it has heterogeneous distribution of  stars in the CMD (Fig. \ref{cmd_1}), strongly suggesting  a  non-physical object.

The statistical method comparing object and field CMDs has provided complementary results.  It tests whether  an object  has  a CMD compatible with that of the field. We conclude that  72 \% of the  studied objects have a probability $p<2\%$  of being representative of the field (Table \ref{table:2}).

Proper motions  are  indicators of the internal kinematics of the  studied objects. Since we are dealing with intrinsically underpopulated stellar systems, it is important to build composite CMDs and proper-motion diagrams. The OCR composite CMDs show similarities with those of individual open clusters. The OCR composite  proper-motion diagrams suggest their classification into two distinct patterns: Group A with one and Group B with two pronounced velocity peaks. The distinction between  binary  and single stars and their mapping in the CMD suggest  evolutionary stages.

Figure \ref{age} shows the age distribution of OCRs and POCRs with $I_f > 50\%$ separated in Groups A and B. The peak of the distribution is in the range 0.5-1.0 Gyr, but older objects also exist. Figure \ref{rgc} shows the objects in Groups A and B   in terms of  age as a function of Galactocentric distance  $R_{gc}$ (left panel) and  height from the plane $|z|$ as a function of  $R_{gc}$   (right panel). For comparison we included 230 open clusters from Chen et al. (\cite{chen}). The Sun is assumed  to be at  8 kpc (Reid \cite{reid}). Despite the small  sample and the error bars, if we consider that the Group B objects tend to be  located inside the Solar circle, this configuration could  suggest that tidal effects related to $R_{\rm{gc}}$ affect the evolution  of remnants (Table \ref{table:3}).  Considering a shorter  distance ($R_{gc}=$ 7.2 kpc)  of the Sun to the Galactic center (Bica et al. \cite{bica06}),  tidal effects would be stronger.

 \begin{table*}
      \caption[]{Final classification for the POCRS}
         \label{table:5}
  \centering
\begin{tabular}{lccccc}
\hline 
Nome& Type & Profile &R$^{2}$ &I$_{f}$ &Q \\
\hline
\hline
NGC\,3680 & L & y & y & y   &OC\\
\hline
\hline
NGC\,6481 & C & y & y & y  & OCR\\

NGC\,6994 & C &  y & n & n  & FF\\

NGC\,6863 & C & y & y & y  &OCR\\

NGC\,1663 & L &y & y & y  &OCR\\

ESO\,425\,SC\,6& L&  y & y & y  &OCR\\

ESO\,425\,SC\,15& L&y& y   & y & OCR\\

Ruprecht\,3& C&y& y & y & OCR\\

ESO\,426\,SC\,26& L& y & y & y& OCR\\

ESO\,429\,SC\,2& L& y & n & y & POCR\\

Ruprecht\,31& C &y & y & y & OCR\\

Waterloo\,6& C & y & n & n  & FF\\

ESO\,570\,SC\,12& L& y& y & y  & OCR\\

NGC\,1252& L& y& y & y  & OCR\\

NGC\,1901 &L& y& y & y   & OCR\\

Object\,1& C& y & y & n& POCR\\

ESO\,132\,SC\,14 & C & y & y  & n & POCR\\

ESO\,383\,SC\,10&S& y & y & y & OCR\\
 
Lyng\aa\,8 &C& y & y & y & OCR\\
\hline   
\end{tabular}
\end{table*}

\section{Concluding remarks}

We studied 18 POCRs by means of 2MASS J and H photometry and UCAC2 proper motions. We analyzed  structural properties and  distributions of stars in CMDs of POCRs  and offset fields  using a statistical method.  Ages were derived, together with reddening values and distances to the Sun, using the CMDs and Padova isochrones. We defined an  index of isochrone fit.  All these methods are useful criteria for remnant characterization. The analysis  indicates the presence of 13 open cluster  remnants in the sample (Table \ref{table:5}). 
\begin{figure}
\resizebox{\hsize}{!}{\includegraphics[width=12cm]{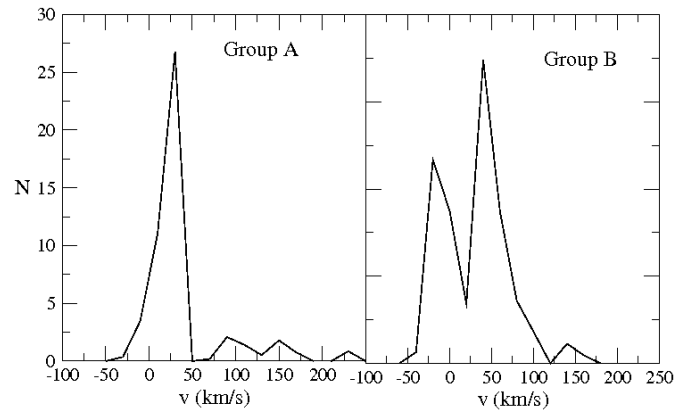}}
\caption{Distribution of  velocities projected on the sky  reduced to the rest velocity. Left panel: Group A. Right panel: Group B. A  velocity bin of 20 km/s was used.}
\label{groupab}%
\end{figure}

\begin{figure*}
  \centering 
  \includegraphics[width=12cm]{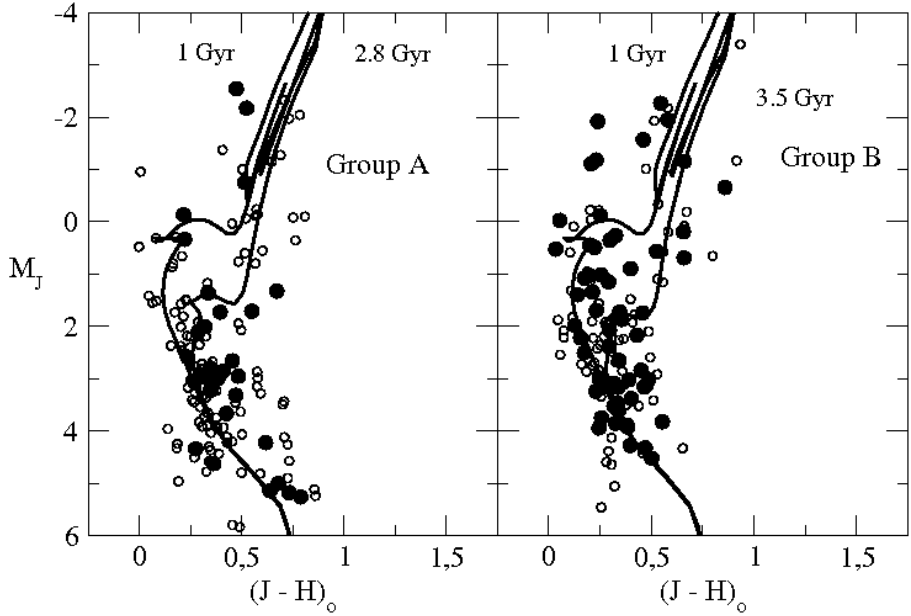}
  \caption{Absolute CMDs for Groups A and B (left and right panels, respectively). Stars with velocities ($v_p -v_M$) higher than 25 km/s (filled circles) are superimposed on the CMDs. Padova isochrones are shown as reference.}
             \label{gpab_cmd}
	     \end{figure*}

Open clusters are known to contain a significant fraction of multiple systems as a consequence of their dynamical evolution. The fraction of multiple
systems increases and tends to concentrate in the central regions, thus
changing the initial spatial distribution of stars (Takahashi \& Portegies Zwart \cite{takahashi}). Numerical simulations show that open cluster remnants are
expected to be rich in binary stars. As an example, de La Fuente Marcos (\cite{fuente1997}) stopped  simulations when the cluster population was  10 stars and the number of the remaining binaries was 3 or 4.  Recent studies comparing observational and theoretical results indicate that cluster remnants can form highly hierarchical high-order multiple systems; and considering that the number of such objects can be larger, cluster remnants may be sufficient to account for all observed higher-order multiple systems (Goodwin \& Kroupa \cite{kroupa}).

The present overall results indicate evolutionary differences among sample objects.  POCRs show two distinct structures, compact and loose, and we verify  a correspondence between the radial stellar density profiles and these structures. Sample objects are basically  consistent with  OCRs that gradually lose stars as compared to open clusters. The composite proper motion diagrams shows evidence of binary stars, as expected from high binary fractions in dynamically evolved systems.  The presence of a sequence of  binaries in the composite CMD, replacing the MS of single-stars in  Group B, suggests the existence of distinct stages along the dynamical evolution of open cluster remnants. In addition, most of the compact objects are included in Group B and tend to be more internally located in the disc. If two distinct structures might be associated to different evolutionary phases, then  transitional objects should be detected. ESO\,570\,SC12 is visually loose (Fig. \ref{FigXDSS}) but was compact in the radial density profile (Fig. \ref{perfis2}). However, a large sample would be necessary to constrain such possible links. For the moment we cannot exclude the possibility that Groups A and B might be descendants of different types of open clusters.

For the   first time a  study presents  a considerable   number of  POCRs analyzed by means of the same methods for the  characterization of their  common properties. We verify that is necessary  to combine different methods in order  to constrain the object nature owing to data incompleteness,  observational uncertainties, and intrinsically low  statistics. The existence of compact and loose POCRs suggests relatively stable remnant structures that  dynamically   survive beyond open clusters.  

We confirm  NGC\,1252, NGC\,1901, Ruprecht\,3, and NGC\,1663 as open cluster remnants, and   NGC\,6994 as a field fluctuation, in agreement with  previous studies (Pavani et al. \cite{pavani2001b}; Pavani et al. \cite{pavani2003}; Baume et al. \cite{baume}; Carraro \cite{carraro2000}; Odenkirchen \& Soubiran \cite{odenkirchen}). Evidence is found that  NGC\,6481, NGC\,6863, ESO\,425\,SC6, ESO\,425\,SC15, ESO\,426\,SC26, ESO\,570\,SC12, Ruprecht\,31, ESO\,383\,SC10, and  Lyng\aa\,8  are also open cluster remnants, while Waterloo\,6 is a field fluctuation.  ESO\,429\,SC2, Object\,1, and ESO\,132\,SC14 are maintained as POCRs for more detailed future studies.

The present work offers observational evidence of the existence in nature of open cluster remnants, as expected from theoretical and numerical studies. As prospective work to answer fundamental questions, we mention (i) would the distinction between compact and loose structures be related to that of the progenitor open cluster and/or the evolutionary processes and (ii), if evolutionary stages in remnants are confirmed, what detailed processes would cause them?

 \begin{figure}
  \centering 
  \includegraphics[width=8cm]{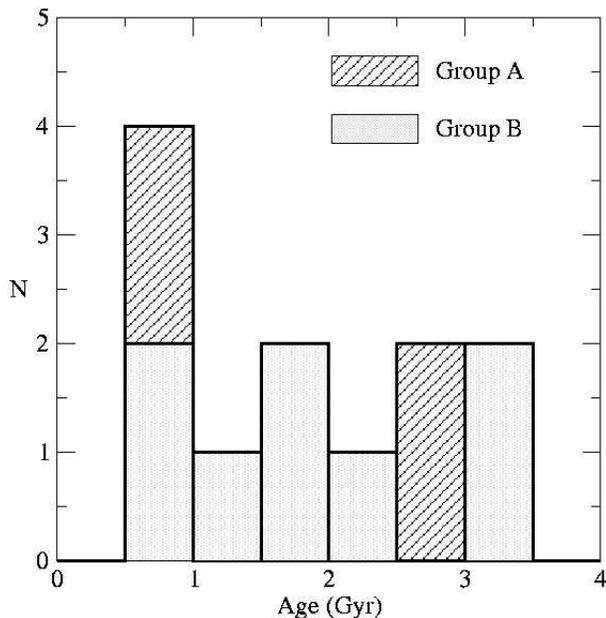}
  \caption{Age histogram for OCRs and POCRs.}
             \label{age}%
	     \end{figure}

\begin{figure*}
  \centering 
  \includegraphics[width=15cm]{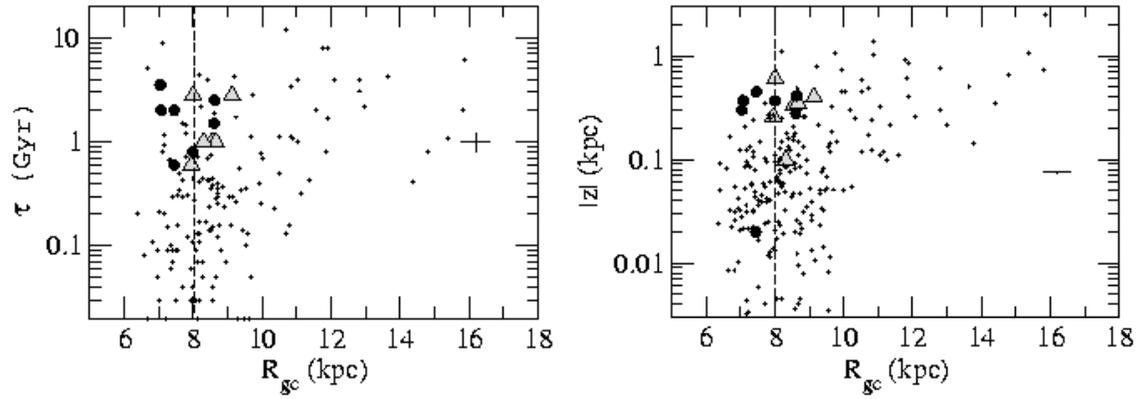}
\caption{Age of OCRs and POCRs vs. Galactocentric distance R$_{gc}$ (left panel), and  height from the plane $|z|$ vs. Galactocentric distance R$_{gc}$ (right panel). The Sun is at 8.0 kpc. We distinguish  Group A (triangles) and Group B objects (filled circles). Open clusters from Chen et al. (\cite{chen}) are shown (plus sign)}.
             \label{rgc}%
	     \end{figure*}

\begin{acknowledgements}
We thank Dr. Leandro Kerber for  support and advice concerning the development of the statistical method (Sect. 3), and Dr. Charles Bonatto for interesting discussions.  
This work makes use of data products from the Two Micron All Sky Survey, which is a joint project of the University of Massachusetts and Infrared Processing and Analysis/California Institute of Technology, funded by the National Aeronautics and Space Administration and the National Science Foundation. We also made use of UCAC2 data. We acknowledge support from the Brazilian Institutions CNPq and FAPESP.
\end{acknowledgements}

\end{document}